\documentclass[sn-mathphys,Numbered]{sn-jnl}

\usepackage[markup=underlined]{changes}
\definechangesauthor[color=red]{MS}
\definechangesauthor[color=blue]{xxx}
\usepackage{todonotes}
\setcommentmarkup{\todo[color={authorcolor!20},size=\scriptsize]{#3: #1}}

\usepackage{graphicx}%
\usepackage{multirow}%
\usepackage{amsmath,amssymb,amsfonts}%
\usepackage{amsthm}%
\usepackage{mathrsfs}%
\usepackage[title]{appendix}%
\usepackage{xcolor}%
\usepackage{textcomp}%
\usepackage{manyfoot}%
\usepackage{siunitx}
\usepackage{booktabs}%
\usepackage{algorithm}%
\usepackage{algorithmicx}%
\usepackage{algpseudocode}%
\usepackage{hyperref}       %
\usepackage[version=4]{mhchem}
\usepackage[utf8]{inputenc} %
\usepackage[T1]{fontenc}    %

\usepackage{url}            %
\usepackage{booktabs}       %
\usepackage{nicefrac}       %
\usepackage{microtype}      %
\usepackage{xcolor}         %
\usepackage[export]{adjustbox}
\usepackage{wrapfig,amsthm,textcomp,amssymb}
\usepackage{colortbl}
\usepackage{pifont}%
\usepackage[capitalise,noabbrev,nameinlink]{cleveref}
\usepackage{adjustbox}

\usepackage{subfigure}
\usepackage[percent]{overpic}
\usepackage{lineno}

\usepackage{soul}
\usepackage{cancel}
\usepackage{graphicx}
\usepackage{listings}%
\definecolor{good}{HTML}{ffd2ad}
\definecolor{good}{HTML}{74c476}

\begin{document}
\title[Article Title]{Iterative Pretraining Framework for Interatomic Potentials
}


\author[1,2]{\fnm{Taoyong} \sur{Cui}}
\equalcont{These authors contributed equally to this work.}
\author[2,3]{\fnm{Zhongyao} \sur{Wang}}
\equalcont{These authors contributed equally to this work.}
\author[2]{\fnm{Dongzhan} \sur{Zhou}}
\author[2]{\fnm{Yuqiang} \sur{Li}}
\author[2]{\fnm{Lei} \sur{Bai}}
\author[1,2]{\fnm{Wanli} \sur{Ouyang}}
\author*[2,4]{\fnm{Mao} \sur{Su}}\email{sumao@pjlab.org.cn}
\author*[2]{\fnm{Shufei} \sur{Zhang}}\email{zhangshufei@pjlab.org.cn}

\affil[1]{\orgname{The  Chinese University of Hong Kong}, \orgaddress{\city{Hongkong}, \postcode{999077}, \country{China}}}

\affil[2]{\orgname{Shanghai Artificial Intelligence Laboratory}, \orgaddress{\city{Shanghai}, \postcode{200232}, \country{China}}}

\affil[3]{\orgdiv{College of Computer Science and Artificial Intelligence}, \orgname{Fudan University}, \orgaddress{\city{Shanghai}, \postcode{200433}, \country{China}}}

\affil[4]{\orgname{Shenzhen Institute of Advanced Technology, Chinese Academy of Sciences}, \orgaddress{\city{Shenzhen}, \postcode{518055}, \country{China}}}







\abstract{Machine learning interatomic potentials (MLIPs) enable efficient molecular dynamics (MD) simulations with ab initio accuracy and have been applied across various domains in physical science. However, their performance often relies on large-scale labeled training data. While existing pretraining strategies can improve model performance, they often suffer from a mismatch between the objectives of pretraining and downstream tasks or rely on extensive labeled datasets and increasingly complex architectures to achieve broad generalization. To address these challenges, we propose Iterative Pretraining for Interatomic Potentials (IPIP), a framework designed to iteratively improve the predictive performance of MLIP models. IPIP incorporates a forgetting mechanism to prevent iterative training from converging to suboptimal local minima. Unlike general-purpose foundation models, which frequently underperform on specialized tasks due to a trade-off between generality and system-specific accuracy, IPIP achieves higher accuracy and efficiency using lightweight architectures. Compared to general-purpose force fields, this approach achieves over 80\% reduction in prediction error and up to 4× speedup in the challenging Mo–S–O system, enabling fast and accurate simulations.

}

\keywords{Machine Learning Interactomic Potentials, Pretraining, Graph Neural Network}



\maketitle

\section{Introduction}\label{sec1}

Molecular dynamics (MD) simulations serve as a tool for revealing the microscopic dynamical behavior of matter and play a key role in areas such as materials design, drug discovery, and analysis of chemical reaction mechanism~\cite{butlerMachineLearningMolecular2018,drorPathwayMechanismDrug2011,zengComplexReactionProcesses2020,karplusMolecularDynamicsSimulations2002,devivoRoleMolecularDynamics2016}. Traditional molecular dynamics~\cite{karplusMolecularDynamicsSimulations1990,10.1145/1654059.1654099,devivoRoleMolecularDynamics2016} employs empirical force fields to describe interatomic interactions. Although these parameterized potentials enable efficient computation, their fixed functional forms struggle to capture complex quantum effects~\cite{BP-limit,ceriotti2016nuclear}. In contrast, ab initio molecular dynamics (AIMD)~\cite{AIMD} can provide more accurate potential energy surfaces using first-principles calculations. However, computational complexity hinders the use of AIMD for large systems and long timescales~\cite{wangInitioCharacterizationProtein2024,beckePerspectiveFiftyYears2014,GordenBell}. This intrinsic trade-off between accuracy and efficiency remains a bottleneck in the advancement of the atomistic simulation techniques \cite{MLFF}.

In recent years, machine learning interatomic potentials (MLIPs)~\cite{behler2007generalized,friederich2021machine, cui2025online,batznerAdvancingMolecularSimulation2023}, leveraging data-driven models to fit the results of first-principles calculations, have emerged as a transformative approach in MD simulations. Compared to traditional empirical potentials, MLIPs offer greater flexibility in capturing complex atomic interactions while achieving an optimal balance between accuracy and computational efficiency ~\cite{friederich2021machine, liaoequiformerv2}. However, a critical challenge remains: the performance of MLIPs relies heavily on access to extensive, high-quality first-principles training data. Furthermore, the coverage and diversity of the training data also affect model performance. Inadequate sampling or biased data distributions can severely compromise reliability, especially for complex systems. The scarcity of high-quality data limits the real-world application of MLIPs in complex materials and molecular systems~\cite{merchantScalingDeepLearning2023}. 

To address the issue of data scarcity, the most straightforward approach is to expand the training dataset through active learning ~\cite{active1,active2,active3,active4,eip}. However, the cost of exploring the configurational space remains prohibitively high. In comparison, pretraining serves as a promising alternative strategy that alleviates the reliance on high-quality labeled data~\cite{dpa2}. Existing pre-training solutions can be broadly categorized into self-supervised ~\cite{gpip,frad} and fully supervised methods ~\cite{uma,mace,dpa2,mattersim}. The former attempt to enhance force field accuracy in low-data regimes by integrating large amounts of unlabeled configurations and applying self-supervised learning techniques, thereby avoiding the need for expensive quantum calculations. Yet, there is often a gap between pretraining objectives and downstream fine-tuning tasks, which can hinder effective model learning ~\cite{zakarias2024bissl}. Fully supervised methods, by contrast, leverage labeled data from first-principles calculations. Still, distributional shifts between pretraining and fine-tuning domains often impair generalization to complex target systems. Recently, foundation models trained using the supervised paradigm have garnered widespread attention. 
However, these models are becoming increasingly overparameterized and architecturally intricate in pursuit of better data fitting and generalization.  In particular, to strictly enforce higher-order Euclidean symmetries, many employ tensor product operations ~\cite{nequip,allegro,mace_fm,sevennet}, which are computationally intensive and significantly degrade inference efficiency. More importantly, foundation models often struggle to balance generalization across diverse tasks with precision on specialized applications. While both fine-tuning and distillation can partially alleviate the aforementioned limitations, fine-tuning does not address the bottleneck of inference efficiency, and distillation may degrade model accuracy due to  overfitting to the teacher’s distinctive representation patterns.

In this work, we propose an Iterative Pretraining framework (IPIP) to address the above mentioned challenges and achieve a synergistic improvement in simulation accuracy and efficiency. Specifically, we design a scalable, iterative optimization-based pretraining strategy that avoids introducing additional quantum calculations, thereby preserving computational efficiency. 
The IPIP framework begins with MD simulations guided by an MLIP foundation model to systematically explore a broad and diverse configurational space. A lightweight student model is then fine-tuned using high-quality data (e.g., DFT calculations). In subsequent iterations, the framework employs this student model for simulations to expand pretraining dataset. Through multi-stage iterative refinement, the model converges toward improved generalization and predictive accuracy across downstream interatomic tasks. To validate the IPIP method, we performed extensive benchmarking across diverse molecular systems, achieving a 20\% improvement in accuracy over existing pretraining approaches. To further assess its robustness, we tested IPIP on the challenging Mo–S–O system, where conventional force fields typically fail. Remarkably, IPIP not only maintains stability under reactive conditions but also delivers a 4× speedup in molecular dynamics simulations while reducing prediction errors by >80\% compared to general-purpose force fields.

\begin{figure}[tbp]  
    \begin{center}  
        \begin{overpic}[width=0.9\textwidth]{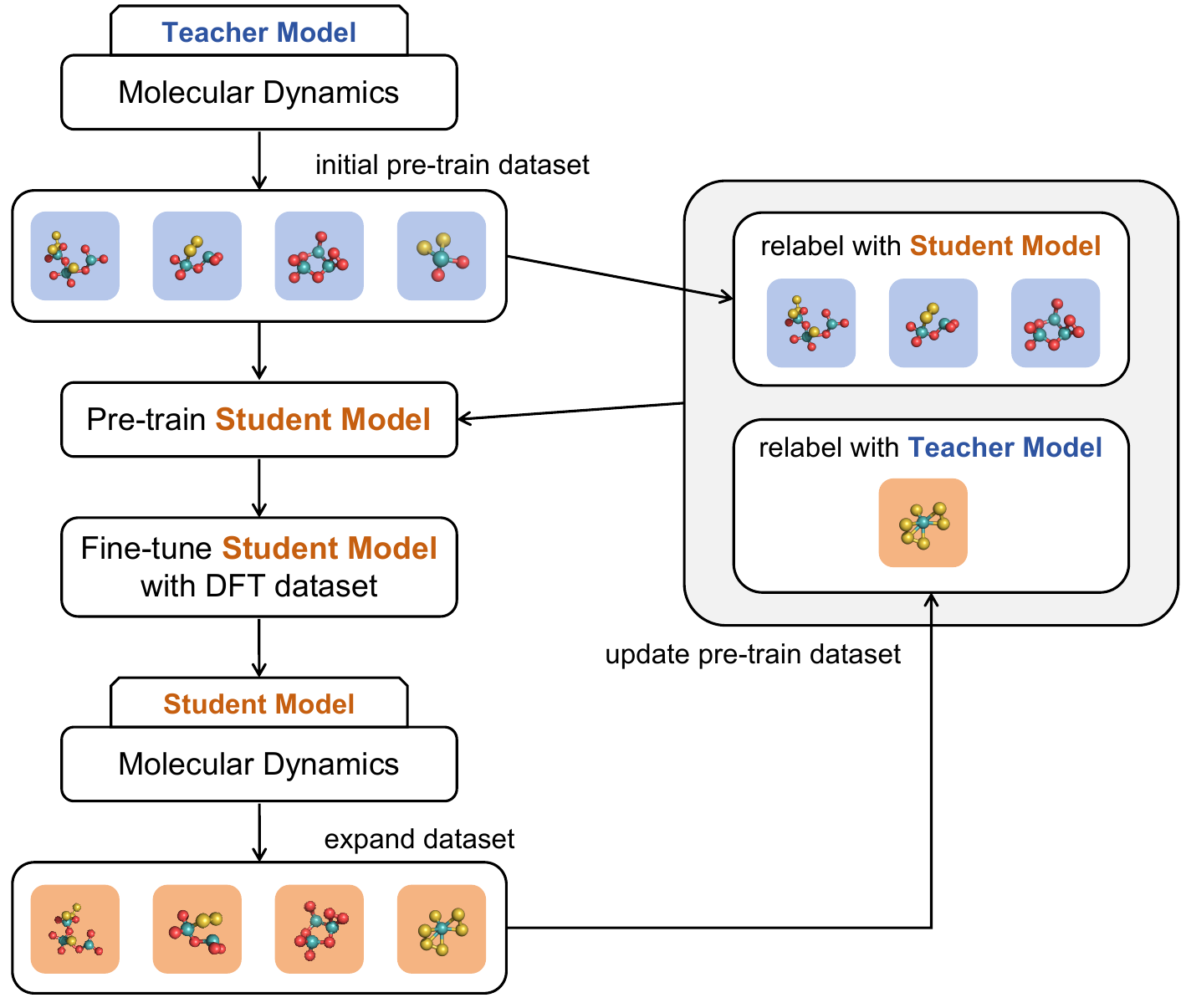}  
        \end{overpic}  
    \end{center}  
    \caption{\textbf{Overview.} IPIP is an iterative enhancement framework that boosts MLIP model predictions. The process consists of three key steps: data preparation, pre-training and fine-tuning, and dataset expansion. This framework develops a iterative MLIPs pre-training framework aimed at balancing accuracy and efficiency through a multi-stage iterative strategy without requiring costly DFT computations.
    }  
    \label{fig:pipeline}  
\end{figure}
\section{Results}\label{sec2}

\subsection{Preliminaries}\label{pre}
MLIPs aim to learn the potential energy surface (PES) of an atomic system from data.
A specific molecular configuration is uniquely described by the set of atomic numbers $Z = \{z_1, z_2, \dots, z_n\} \in \mathbb{Z}^{+}$ and their corresponding 3D coordinates $R = \{r_1, r_2, \dots, r_n\} \in \mathbb{R}^{n \times 3}$.
The core task of an MLIP is to learn a mapping from the atomic structure $(Z, R)$ to the total potential energy $E \in \mathbb{R}$. The forces acting on each atom $i$ are then derived as the negative gradient of the energy with respect to its position:
\begin{equation}
    F_i = -\nabla_{r_i} E(Z, R).
\end{equation}

This work focuses on Graph Neural Network (GNN) based MLIPs. In these models, each atom's feature vector $x_i^0$ is initialized from its atomic number $z_i$ via a learnable embedding, $x_i^0 = f_{\text{emb}}(z_i)$. These features are then iteratively refined over $L$ message-passing layers where each atom aggregates information from its neighbors. Finally, the total potential energy $\hat{E}$ is computed by summing atomic energy contributions, where each contribution $E_i$ is decoded from the final atom feature $x_i^L$ by a multi-layer perceptron (MLP):
\begin{equation}
    \hat{E} = \sum_{i=1}^{n} \text{MLP}(x_i^L). \label{eq:readout}
\end{equation}

\subsection{IPIP framework}\label{framework}
This study develops a scalable pretraining framework for MLIPs named IPIP to balance accuracy and efficiency. IPIP is an iterative enhancement pretraining framework designed to improve the predictive capability of MLIP models. The overall workflow consists of the following four key steps:

\textbf{Pretraining Data Preparation.} We begin by using an existing universal potential as a teacher model to perform molecular dynamics simulations starting from multiple initial configurations. These simulations generate an initial pretraining dataset capturing rich atomic-structural configurations and corresponding pseudo labels (e.g., energies and forces).

\textbf{Pretraining and Fine-tuning} A student model (e.g., ViSNet or PaiNN), characterized by higher inference efficiency and lower computational cost, is then trained on the initial pretraining dataset. This step enables the student model to acquire a preliminary generalization ability across diverse configuration spaces. Next, the student model is fine-tuned using a small number of high-accuracy DFT-labeled samples from the target task. This calibration step corrects prediction errors in critical structural regions and significantly improves the overall accuracy of the model.

\textbf{Iterative Pretraining Dataset Construction.} Unlike conventional one-shot pretraining approaches, IPIP introduces a closed-loop iterative mechanism that continuously refines the pretraining dataset using feedback from MLIP-driven MD simulations. This strategy allows the model to not only learn from initial datasets but also actively explore more diverse regions in the configuration space through MD simulations. In each iteration, the current student model is employed to perform large-scale MD simulations, automatically sampling a broader and more diverse set of atomic configurations. Particular emphasis is placed on collecting edge conformations—structures that are more likely to induce instability or prediction failures during simulation. These challenging cases often represent underexplored regions in the model’s representation space and are critical for improving robustness. 

Pseudo-labels are generated through two complementary strategies and combined to construct an updated pretraining dataset:

\begin{itemize}
    \item \textbf{Pseudo-labeling}: To leverage the knowledge learned from DFT data and propagate it to exsiting pretraining data, 10\% of the original pretraining data is randomly discarded, and the student model generates new pseudo-labels (energy and force) for the remaining data.
    \item \textbf{Foundation Model Re-annotation}: To correct potential errors from the student model during simulations, an MLFF foundation model is used to generate pseudo-labels for newly sampled configurations obtained from student model-driven simulations.
\end{itemize}

The updated pretraining dataset is then used to retrain the student model. Through multiple iterations, this self-enhancing training loop progressively improves the model's accuracy, stability, and transferability across diverse atomic environments.

\subsection{Experiments}\label{Experiments}

\begin{figure}[t]
    \centering
    \includegraphics[width=0.8\linewidth]{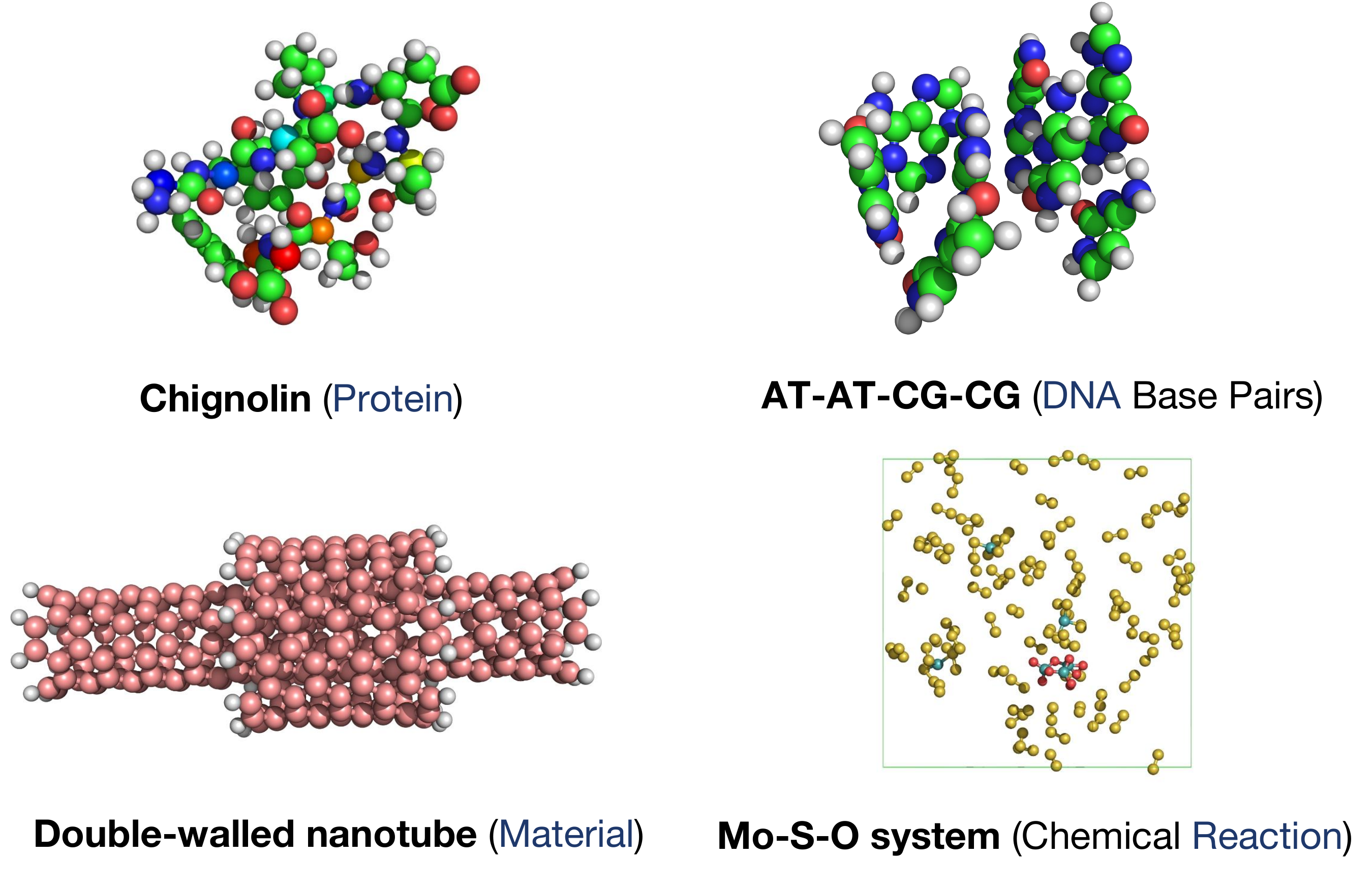}
    
    \caption{\textbf{Representative atomic structures of diverse systems.} 
    (a) Chignolin, a short biomolecule. (b) AT-AT-CG-CG, a biomolecular system from the MD22 dataset. 
    (c) Double-walled carbon nanotube (MD22 dataset), representative of low-dimensional materials. 
    (d) \ce{MoS6}, a cluster formed during the sulfidation of \ce{Mo3O9}.}
    
    \label{fig:2-2}
    \vspace{-3.0ex}
\end{figure}
We conducted a series of experiments across multiple datasets to evaluate the performance of IPIP within widely used ViSNet~ and PaiNN~. Both models adopt first-order equivariant architectures, offering a favorable balance between computational efficiency and predictive accuracy. 

We first benchmarked IPIP on the MD-22 dataset~\cite{md22}, demonstrating clear improvements over state-of-the-art pretraining methods and knowledge distillation. To further assess its applicability in realistic chemical scenarios, we extended our evaluation to two additional systems: the folding dynamics of Chignolin~\cite{chignolin}, a fully atomistic peptide model representative of biomolecular microfolding, and the high-temperature sulfidation of \ce{Mo3O9} clusters into \ce{MoS6}/\ce{MoS7} species in sulfur vapor. The Chignolin system tests IPIP’s ability to stably model complex biomolecular dynamics, while the \ce{Mo3O9} sulfidation captures a prototypical non-equilibrium chemical transformation involving bond breaking, structural rearrangement, and the formation of new coordination motifs.

The IPIP framework is evaluated using two powerful GNN backbones:
\begin{itemize}
    \item \textbf{PaiNN} ~\cite{painn} is an equivariant GNN that preserves E(3) symmetry by modeling directional information through vectorial messages.
    \item \textbf{ViSNet} ~\cite{visnet} is a high-performing equivariant architecture that uses runtime geometry computation (RGC) for low-cost modeling of many-body interactions.
\end{itemize}

For re-annotation in our pre-training loop, we leverage a state-of-the-art foundation model as a teacher model:
\begin{itemize}
    \item \textbf{MACE-OFF}~\cite{mace} is a highly accurate universal potential that achieves its performance through higher-order equivariant message passing.
\end{itemize}

\begin{table*}[t]
\centering
 \caption{Mean absolute errors (MAE) of energy (kcal) and force (kcal/mol/\AA) for various molecular systems using different pretraining methods on ViSNet model. The best result in each row is highlighted in bold.}
\vspace{-2mm}
\begin{threeparttable}
{\footnotesize  
\setlength{\tabcolsep}{4.5pt}  
\renewcommand{\arraystretch}{1.15}  

\begin{tabular}{lcccccc}
\toprule
System &  & Baseline & GPIP~\cite{gpip} & Frad~\cite{frad} & Ours \\ \midrule

\multirow{2}{*}{Ac-Ala3-NHMe} 
  & Energy & 0.0896 & 0.0867 & 0.0744 & \textbf{0.0572} \\
  & Force  & 0.0801 & \textbf{0.0729} & 0.0740 & \textbf{0.0729} \\ \midrule

\multirow{2}{*}{DHA} 
  & Energy & 0.0938 & 0.0899 & 0.0867 & \textbf{0.0801} \\
  & Force  & 0.0618 & 0.0622 & 0.0612 & \textbf{0.0542} \\ \midrule

\multirow{2}{*}{Stachyose} 
  & Energy & 0.1351 & 0.1223 & 0.1289 & \textbf{0.0911} \\
  & Force  & 0.0934 & 0.0887 & 0.0924 & \textbf{0.0694} \\ \midrule

\multirow{2}{*}{AT-AT} 
  & Energy & 0.1311 & 0.1156 & 0.1187 & \textbf{0.0889} \\
  & Force  & 0.0959 & 0.8934 & 0.8864 & \textbf{0.0792} \\ \midrule

\multirow{2}{*}{AT-AT-CG-CG} 
  & Energy & 0.2030 & 0.1842 & 0.1821 & \textbf{0.1601} \\
  & Force  & 0.1289 & 0.1154 & 0.1157 & \textbf{0.1088} \\ \midrule

\multirow{2}{*}{Buckyball catcher} 
  & Energy & 0.5001 & 0.4564 & 0.3902 & \textbf{0.2727} \\
  & Force  & 0.1843 & 0.1581 & 0.1657 & \textbf{0.1054} \\ \midrule

\multirow{2}{*}{Double-walled nanotube} 
  & Energy & 0.8003 & 0.6786 & 0.6247 & \textbf{0.5421} \\
  & Force  & 0.3627 & 0.3764 & 0.3478 & \textbf{0.2344} \\ \midrule

\end{tabular}
}
\end{threeparttable}
\label{table:main1}
\vspace{-3mm}
\end{table*}

\begin{table*}[t]
\centering
\vspace{-3mm}
\caption{Mean absolute errors (MAE) of energy (kcal/mol) and force (kcal/mol/\AA) for various molecular systems using different model distillation methods on ViSNet model. The best result in each row is highlighted in bold.}
\vspace{-2mm}
\begin{threeparttable}
{\footnotesize
\setlength{\tabcolsep}{5pt}
\renewcommand{\arraystretch}{1.15}

\begin{tabular}{lcccccc}
\toprule
System &  & teacher model~\cite{mace} & n2n~\cite{n2n} & a2a ~\cite{hessian} & Hessian~\cite{hessian} & Ours \\ \midrule

\multirow{2}{*}{\textbf{Buckyball catcher}} 
  & Energy & -      & 0.4733 & 0.4752 & 0.4032 & \textbf{0.2727} \\
  & Force  & 1.6554 & 0.1787 & 0.1797 & 0.1548 & \textbf{0.1054} \\ \midrule

\multirow{2}{*}{\textbf{Double-walled nanotube}} 
  & Energy & -      & 0.7457 & 0.7687 & 0.6724 & \textbf{0.5421} \\
  & Force  & 1.1065 & 0.3411 & 0.3424 & 0.2947 & \textbf{0.2344} \\ \midrule

\end{tabular}

}
\end{threeparttable}
\label{table:variants}
\vspace{-3mm}
\end{table*}
\textbf{MD-22.} The MD22 dataset encompasses a diverse range of realistic and structurally complex systems, including proteins, lipids, carbohydrates, nucleic acids, and supramolecular assemblies. Table 1 summarizes the mean absolute errors (MAEs) for energy and force predictions with ViSNet using two different state-of-the-art self-supervised pretraining strategies (GPIP ~\cite{gpip} and Frad~\cite{frad}). The results show that the IPIP-based model outperforms all compared methods across all seven molecular systems, demonstrating strong generalization capability in complex chemical environments. In addition, we conducted further comparisons on the two largest and most structurally complex systems in the dataset, by evaluating IPIP against recent state-of-the-art distillation methods (n2n, a2a, and Hessian distillation schemes) as well as a universal potential (MACE-OFF). As shown in Table 2, IPIP achieves superior predictive accuracy compared to the universal potential, while also exhibiting better generalization than other distillation based approaches.

\begin{figure}[t]
    \centering
    \includegraphics[width=1.0\linewidth]{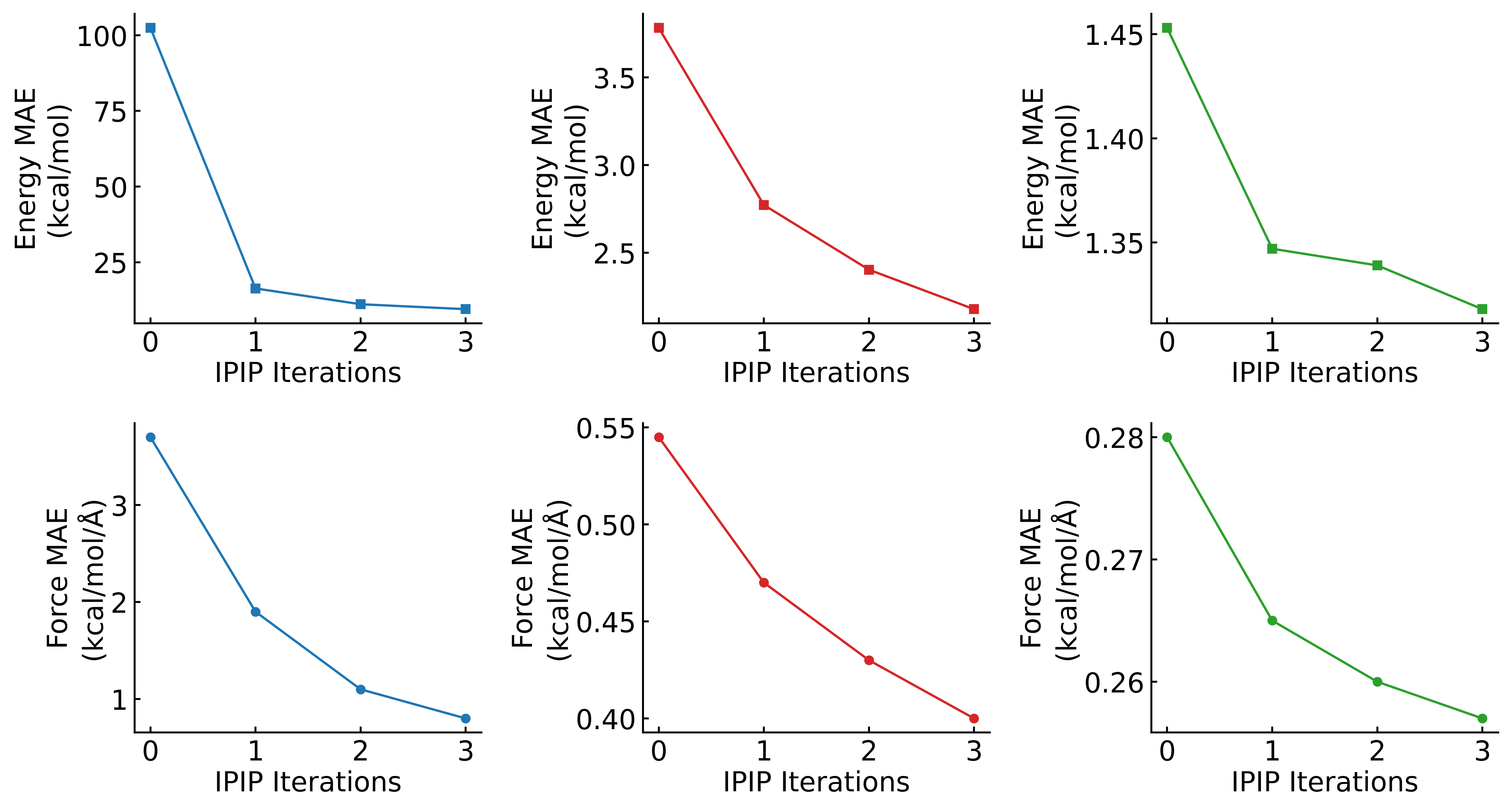}
    
    \caption{\textbf{IPIP experimental results on the Chignolin dataset using the ViSNet model.} IPIP results after three iterations on the Chignolin dataset using different training set sizes. The blue, red, and green curves correspond to 20k, 100k, and 600k training samples, respectively.}
    \label{fig:2-2}
    \vspace{-3.0ex}
\end{figure}
\textbf{Chignolin.}To evaluate the practical applicability of IPIP in real-world scenarios, we selected the mini-protein Chignolin, consisting of 166 atoms, as a test case. This dataset contains approximately 2,000,000 configurations, annotated with corresponding energies and forces computed at the DFT level. We conducted a systematic evaluation of IPIP across varying dataset sizes, ranging from 20,000 to 600,000 configurations. The results demonstrate that our method maintains strong generalization performance under different data availability conditions. Furthermore, we investigated the prediction performance of IPIP across multiple training iterations. The results show that each iteration consistently leads to stable and significant improvements in accuracy, highlighting the critical role of the proposed iterative strategy in enhancing model robustness.

\textbf{Mo-S-O system.}
To assess the practical performance of our method, we first tested the baseline model PaiNN trained on the same DFT dataset, chosen for its relatively fast inference speed and competitive accuracy. This baseline model exhibited poor stability, failing to complete the complex reaction between \ce{Mo3O9} and \ce{S2} in 13\% of 100 independent simulations initialized with different random seeds, thus unable to reliably capture the reaction pathway involving ring-opening and decomposition into smaller sulfided clusters. In contrast, our model trained on the identical dataset demonstrated significantly improved robustness and stability, consistently reproducing the full reaction process across all simulation trials.  This indicates that our training strategy effectively enhances the model’s ability to handle challenging reactive dynamics. Analyzing the simulation trajectories, we observed that the ring-opening step proceeds smoothly, followed by a sequential breakdown into smaller sulfided clusters, in agreement with expected chemical behavior.  These results confirm that our approach not only improves stability but also faithfully captures the underlying reaction mechanisms.

\section{Discussion}\label{discusion}

This study introduces IPIP, an iterative pretraining framework that addresses key challenges in developing accurate and efficient machine learning interatomic potentials (MLIPs). By enabling large-scale, diversity-enriched data generation without the need for additional quantum mechanical calculations, IPIP offers a scalable and cost-effective alternative to conventional pretraining methods. A central strength of the framework lies in its closed-loop design, which progressively augments the training set through model-driven sampling and pseudo-labeling. This iterative refinement leads to more stable simulation and more reliable energy and force predictions, particularly for long-timescale molecular dynamics simulations.

Despite the promising results demonstrated here, there remains substantial potential for further development and application of the IPIP framework. Future work could explore integration of advanced uncertainty quantification techniques to more effectively guide data selection and enhance model reliability during iterative training. The method may also be extended to handle more complex chemical systems involving long-range interactions, charged species, and explicit periodic boundary conditions, thereby broadening its applicability across diverse materials and molecular simulations. Moreover, coupling IPIP with active learning strategies or multi-fidelity datasets could accelerate convergence and reduce computational cost. These directions promise to unlock new opportunities for scalable and robust machine learning potentials in materials science and chemistry.

\section{Methods}\label{methods}

\subsection{Dataset settings}

\textbf{MD22 dataset.} MD22 is a recently introduced molecular dynamics trajectory dataset comprising medium- to large-sized molecular systems, ranging from 42 to 370 atoms, and presents considerable modeling challenges. For training models on the  MD22 dataset, we use the same number of molecules as in ~\cite{visnet} for training and validation, and the rest as the test set.  

\textbf{Chignolin dataset.} The Chignolin dataset contains approximately 2,000,000 distinct configurations of this 166-atom mini-protein. As a result, both the training and test sets encompass a broad range of conformational states. We selected subsets comprising 1\%, 5\%, and 30\% of the dataset for training and validation, while using the remaining configurations for testing.

\begin{figure}[t]
    \centering
    \includegraphics[width=1.0\linewidth]{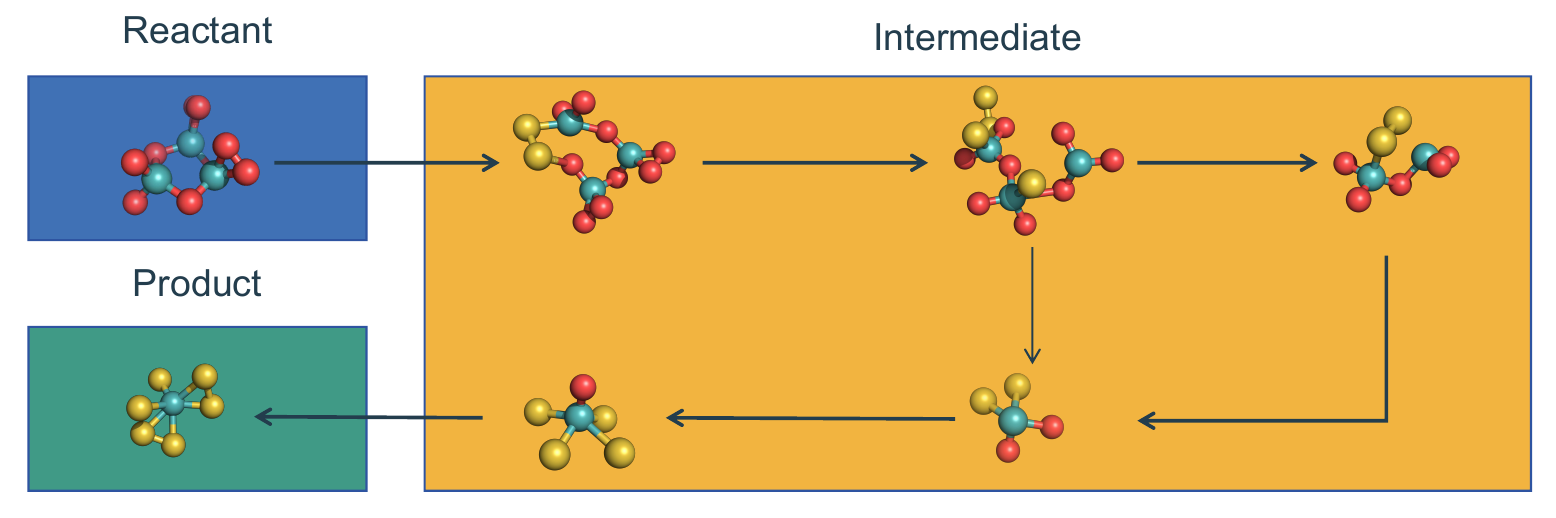}
    
    \caption{Reaction mechanism for the sulfidation of \ce{Mo3O9}. The process involves the initial insertion of \ce{S2} into the \ce{Mo3O9} ring, leading to ring-opening and the formation of a chain intermediate. The chain then breaks down into three separate Mo-included clusters that are fully sulfided to form the final \ce{MoS6} product. Atoms are colored as follows: Mo (teal), O (red), and S (yellow).}
    \label{fig:2-2}
    \vspace{-3.0ex}
\end{figure}
\textbf{Mo-S-O system.}
A comprehensive first-principles dataset for the Mo-S-O system was constructed specifically for this work, encompassing approximately 65,000 configurations. The dataset's structural diversity was achieved through a dual generation methodology: periodic systems were created via an active learning workflow combining ab initio molecular dynamics (AIMD), while aperiodic structures were sampled from a Boltzmann distribution of potential energy surfaces explored by the Stochastic Surface Walking (SSW)~\cite{SSW2013}algorithm, with energies evaluated by the MACE machine learning potential. All reference energies and forces were calculated with the CP2K software package~\cite{cp2k} using the PBE functional~\cite{PBE1996,PBE1997} with D3-BJ dispersion correction~\cite{d32010,d3bj} and GTH pseudopotentials~\cite{GTH1996,GTH1998,GTH2005} . To ensure accuracy and minimize basis set superposition error (BSSE)~\cite{bsse1999}\cite{bsse2007}, the TZV2P-MOLOPT-PBE-GTH basis set~\cite{bsse2007} was employed in conjunction with a high plane-wave cutoff of 850 Ry. For periodic systems, k-point grids were chosen to ensure the product of the sampling density and lattice vector length exceeded 30 {\AA}, whereas aperiodic systems were treated at the Gamma point only. All computations were performed without considering spin polarization.

\subsection{Experimental settings} 

All experiments are implemented by PyTorch 2.1.0+cu118 with four NVIDIA A800 cards. We construct the loss function to the total energy $E$ as well as the forces $F^{j}$ for each atom.

To ensure a fair comparison with ~\cite{visnet}, we follow the original implementation of \textbf{ViSNet} and use the same loss function:
\begin{equation}
\mathcal{L}_{\text{ViSNet}} = \frac{\alpha}{N} \sum_{i=1}^{N} \left( E^{(i)} - \hat{E}^{(i)} \right)^2 
+ \frac{\beta}{n} \sum_{j=1}^{n} \left\| \mathbf{F}^{(j)} - \hat{\mathbf{F}}^{(j)} \right\|^2,
\end{equation}
where $N$ is the number of energy samples, $n$ is the number of force samples (typically the number of atoms), and $\alpha$, $\beta$ are weights that balance the energy and force terms. $\mathbf{F}^{(j)}$ and $\hat{\mathbf{F}}^{(j)}$ are the ground-truth and predicted force vectors for atom $j$, respectively.

For \textbf{PaiNN}, the total loss is defined as:
\begin{equation}
\mathcal{L}_{\text{PaiNN}} = \frac{1}{N} \sum_{i=1}^{N} \left| E^{(i)} - \hat{E}^{(i)} \right| 
+ \frac{\lambda}{n} \sum_{j=1}^{n} \left\| \mathbf{F}^{(j)} - \hat{\mathbf{F}}^{(j)} \right\|,
\end{equation}
where $\lambda$ is a weight balancing the energy and force terms.

\subsection{MD simulations settings}
All molecular dynamics (MD) simulations were performed using the Atomic Simulation Environment (ASE)~\cite{ase} Python library.

\subsubsection{Generation of the initial pretraining dataset}
\textbf{MD22.} To generate a diverse set of configurations for common molecular systems, we utilized the MD22 dataset. For each of the seven substances within the dataset's subset, we sampled 100 structures to serve as initial configurations for MD simulations. Each structure was simulated for 50,000 steps with a 1 fs timestep (50 ps total) using a Langevin thermostat at 300~K. Configurations were collected every 50 steps, with the process repeated until 100,000 configurations were generated for each substance. These simulations were driven by the MACE-OFF24 foundation model. 

\textbf{Chignolin.} For the Chignolin mini-protein, 150 initial configurations were randomly sampled from the original dataset. Each configuration was used to initiate an independent MD simulation under a Langevin thermostat ~\cite{lang1982,lang1983,allen2017computer} at 340~K for 50,000 steps with a 1 fs timestep. A snapshot of the trajectory was saved every 50 steps. After discarding simulations that terminated prematurely due to numerical instabilities, a total of nearly 150,000 stable configurations were collected. The MACE-OFF24 model served as the driving potential.

\textbf{Mo–S–O system.} To generate data for the reactive Mo–S–O system, an initial configuration was prepared consisting of one \ce{Mo3O9} molecule and 72 \ce{S2} molecules within a 55 {\AA} cubic cell under periodic boundary conditions. A set of 50 independent MD simulations was launched, each initiated with a different random seed. Simulations were conducted in the NVT ensemble using a Nosé–Hoover thermostat~\cite{nh1984,nh1985,nh1991} at 1200~K for a duration of 200,000 fs (200 ps). From each trajectory, 2,000 configurations were sampled, yielding a total dataset of 100,000 structures. For this system, a MACE model pretrained on the OMAT24~\cite{omat24} dataset was used to propagate the dynamics.

\subsubsection{Iterative Pretraining Dataset Construction}
Subsequently, the student models (PaiNN or ViSNet) were used to perform MD simulations starting from the same initial configurations as those used in the MACE-driven simulations, under identical simulation settings. Configurations from trajectories that exhibited instabilities were preferentially collected to construct the next-round pretraining dataset, forming an iterative data refinement process.

\subsection{Mo–S–O System Simulations}
\textbf{Stability Analysis.}
To evaluate the stability enhancement provided by IPIP strategy, we compared the performance of models on challenging MD simulations. An initial configuration was generated using the \texttt{packmol} software package, containing a single Mo\textsubscript{3}O\textsubscript{9} molecule and 500 S\textsubscript{2} molecules. The system was prepared at a high density of approximately 0.2 g/cm\textsuperscript{3} to accelerate the exploration of diverse local chemical environments. The initial unit cell was then expanded into a 2$\times$2$\times$2 supercell. For each model being tested, multiple independent simulations were initiated with different random seeds. Each run began with an energy minimization step, followed by an MD simulation in the NVT ensemble using a Langevin thermostat with 2500~K. The simulations were run for 50,000 steps with a timestep of 1~fs (50~ps total). Stability was quantified by the proportion of simulations that terminated prematurely due to numerical instabilities arising from encounters with out-of-distribution (OOD) configurations.

\textbf{Reaction Dynamics Simulation.}
To investigate the chemical reaction pathways, a long-timescale MD simulation was performed. The initial system setup was similar to that used in the stability analysis. The simulation was conducted in the NVT ensemble using a Nosé-Hoover thermostat at a high temperature of 1500~K to access reactive events within a computationally feasible timescale. The system was simulated for 5,000,000 steps with a timestep of 1~fs, totaling 5~ns of simulation time. The resulting trajectory was subsequently analyzed to track the evolution of cluster compositions, with a particular focus on identifying changes in species containing molybdenum atoms.


\bibliography{sn-bibliography}

\end{document}